# Evaluation of the Number of Different Genomes on Medium and Identification of Known Genomes Using Composition Spectra Approach

**Valery Kirzhner**[*1] **& Zeev Volkovich**[2]

[1] Institute of Evolution, University of Haifa, Haifa 31905, Israel;
[2] Software Engineering Department, ORT Braude College of Engineering, Karmiel 21982, Israel.

## Abstract

The article presents the theoretical foundations of the algorithm for calculating the number of different genomes in the medium under study and of two algorithms for determining the presence of a particular (known) genome in this medium. The approach is based on the analysis of the compositional spectra of subsequently sequenced samples of the medium. The theoretical estimations required for the implementation of the algorithms are obtained.

## 1. Introduction

In this paper, we present a method of counting the number of bacterial genomes in the test medium, and also a method to check if a given bacterium is present in the medium. In some special cases, this method also allows determining the multiplicity of each bacterium content in the metagenome. The suggested approach is based on the compositional spectra (CS) method, proposed long time ago [1-5] for the comparison of genomes and/or long genome fragments.

By definition [5], the *compositional spectrum* is the frequency distribution of oligonucleotides of length $l$ (in the literature, referred to as words, $l$-grams, or $l$-mers) which occur in the genome sequence. The existing versions of the method differ mainly in the choice of the set of oligonucleotides, called *support* (*dictionary*), which the frequency distribution (CS) is evaluated for. At present, there exists a large body of research on genome comparisons, which employ different versions of this method and produce results indicative of its validity (see, e.g., [6, 7]).

---

[*] Corresponding author, e-mail: valery@research.haifa.ac.il

Here the CS method is employed to analyze the result of metagenome sequencing, which represents a set of all words of fixed length composing the genomes of the metagenome, with regard to their multiplicity. Since the orientation of each word is unknown, we symmetrize this set by duplicating each word in two possible orientations. Then the set of the metagenome words can be represented as the union of $CS^+$ and $CS^-$ spectra of each genome (with regard to its multiplicity), depending on the chosen sequence direction ($3' \to 5'$ or $5' \to 3'$, respectively). Sum $CS^+ + CS^-$ is referred to as the *barcode spectrum* (*BS*) (see [8]). Below *BS* will be considered as a vector, each coordinate of the vector corresponding to a word from the chosen dictionary. Thus the dimension of vector $N$ grows with the growth of the number of words in the dictionary.

The calculations are based on the following statement, formulated previously [9]: "If the number of the genomes under consideration, $n$, is less than the space dimension, $N$ ($n < N$), *there are no biologically significant reasons for the CS vector of one genome to be in the linear span of the CS vectors of the set of some other genomes*".[1] The same is true for the barcode spectrum. Although the statement of linear independence is empirical, it has been checked for a huge number of genomes [9, 10] and appears to have general character. For more detailed description of the definitions given above see [9].

In Section 2, the concept of a *sample* is introduced as a metagenome extracted from the medium under study. It is essential that, due to the random character of the samples, they, generally speaking, contain a different number of genome copies (have different multiplicity). It is shown that the number of different genomes in the test medium can be calculated by sequentially sampling the medium.

In Section 3, the occurrence of the fixed bacteria in the medium is determined, even in the situation of the presence of unknown genomes in the medium.

---

[1] If $n > N$, then, for purely formal reasons, $n$-$N$ spectra are linear combinations of all the others.

## 2. Calculation of the number of different genomes in the medium

### 2.1. The set of the samples of the medium

We define a *sample* of the medium as a mixture of *n* genomes from the medium, which have different multiplicities in the sample. Consider the properties of sequence $\mathcal{P} = \{p_1, p_2,...\}$ of independent samples of the medium (where $p_i$ is the $i^{th}$ sample). Let $T_i$ be the *vector of genome multiplicities (VGM)* in the $i^{th}$ sample, $T_i = (t_{i1}, t_{i2},...,t_{in})$, where $t_{ij}$ is the number of copies of the $j^{th}$ genome in the $i^{th}$ sample. For the further consideration, it is essential that the number of genome copies in different samples is, generally speaking, different, the character of these quantitative differences depending on the properties of the test medium and on the relative volume of the samples.

Below we present two most probable models of genome occurrences in the sample in terms of VGM.

<u>*Model 1.*</u> Let each sample of unit volume of the medium contain *n* genomes with average multiplicities $m_i$ ($i = 1, 2, ..., n$). The medium is supposed to be spatially homogeneous and time-invariant. Then, the number of copies of the $j^{th}$ genome in each sample can take any value out of set $\{m_j \pm \mu_j\}$ ($\mu_j = 0, 1, 2,..., s-1$), where it is naturally supposed that *s* is much less than $m_j$. Suppose further that each genome multiplicity in the sample is independent of the multiplicities of other genomes in the same sample. Then VGM vectors take the values out of set

$$T_{\mu_1\mu_2...\mu_n} = (m_1 \pm \mu_1, m_2 \pm \mu_2, ..., m_n \pm \mu_n), \quad \mu < s. \quad , \tag{1}$$

where multi-index $\mu_1\mu_2...\mu_n$ determines each vector of set (1). The values of $\mu_*$ are independent of each other and take the values out of set $\{0, 1, 2,..., s-1\}$.

<u>*Model 2.*</u> In the general case, it can be supposed that the $i^{th}$ coordinate of a multiplicity vector $T$, being independent of the other coordinates, can take values out of a finite set of non-negative integers, $a_i = \{a_{ij}\}$, $j=1,..., s$. Then the VGM set consists of all vectors of the type

$$M = \{a_1\} \otimes \{a_2\} \otimes ... \otimes \{a_n\}, \tag{2}$$

where ⊗ designates direct product and

$$T_{j_1 j_2 \ldots j_n} = \{a_{1 j_1}\} \otimes \{a_{2 j_2}\} \otimes \ldots \otimes \{a_{n j_n}\} \ . \tag{3}$$

## 2.2. Basic procedure

The barcode spectrum of each sample is a vector in N-dimensional vector space, which is a linear combination of BS spectra of some genomes of the medium. Thus the number of samples that have linearly-independent spectra is equal to the number of different genomes (i.e., without regard for their multiplicities) in the medium, which was denoted above by $n$.

Now let us sequentially take samples out of the medium and use them for building the basis of the sample BS spectra. This basis (the basic set), certainly, contains the spectrum of the first sample. The spectrum of each consequent sample belongs to the basic set if it is linearly independent of the spectra of already chosen samples. In this way, by sequentially taking the medium samples, it is possible to obtain the full basis for the medium so that none of the further taken samples gives a new independent spectrum. The size of the obtained basis is equal to the number of genomes with linearly independent spectra that are present in the medium.

## 2.3. Statistical estimations for the procedure of building the full sample basis

If vectors are sequentially randomly taken out of an $n$-dimensional space with a standard measure, the $n$ first chosen vectors constitute the space basic with probability 1. Indeed, if the first $p$ vectors chosen in this way are linearly independent, then their linear span is a set of zero measure in the whole space. Therefore, the probability of choosing the next vector from this set is zero. However, if the vectors are to be chosen from any set of special type, the probability of randomly choosing the next vector which is linearly dependent of the already chosen ones can be different from zero.

## 2.4. Estimation of the fractions of the sample set inside and outside the plane

Consider the general scheme described in Model 2.

**Proposition 1**. In *n*–dimensional space, the fraction of the points of set *M* (defined by (2)) which lie outside any plane $L_p$ of codimension *p* ($0 < p < n$) is not less than $1 - s^{-p}$.

**Proof**. Let us prove Proposition by the method of induction on the space dimension, $t$. Namely, the plane of codimension *p* appears when the space dimension *t=p*, i.e., in *p*-dimensional space, and has zero dimension, thus being a point in space *t=p*. In this space, set *M* consists of $s^t$ points $M_{(t=p)} = \{a_{1j}\} \otimes \{a_{2j}\} \otimes ... \otimes \{a_{tj}\}$, not more than one of the points lying on the plane of zero dimension. Consequently, the fraction of the points which lie outside this plane is $(s^t - 1)/s^t$, which equals $1 - s^{-p}$ at *t=p*, as stated in **Proposition**. Now, let **Proposition** be true for some dimension *t*. Let us show that it is also true for dimension *t+1*. In the space of dimension *t+1*, set *M* consists of points $M_{t+1} = \{a_{1j}\} \otimes \{a_{2j}\} \otimes ... \otimes \{a_{t+1,j}\}$. Present this set in the form of union of the planes:

$$M_{t+1} = \{a_{1j}\} \otimes \{a_{2j}\} \otimes ... a_{t+1,1} + \{a_{1j}\} \otimes \{a_{2j}\} \otimes ... a_{t+1,2} + ... + \{a_{1j}\} \otimes \{a_{2j}\} \otimes ... a_{t+1,s},$$

expanding $M_{t+1}$ over the last coordinate. Planes

$$T_r = \{a_{1j}\} \otimes \{a_{2j}\} \otimes ... a_{p+1,r} \quad (r = 1, 2, ..., s) \quad (4)$$

in this representation are, obviously, *t*-dimensional, the dimension of plane $L_p$ in the space of dimension *t + 1* being equal to *t+1-p*. There exist two main cases of the dispositions of planes $T_r$ and $L_p$ with respect to one another.

(1) At *p =1*, one of the planes $T_r$ may coincide with plane $L_p$, whose dimension, in this case, is also *t*. However, the vertexes of all the other planes (4) will lie outside plane $L_p$. Since, in *t+1*-dimensional space, the total number of $T_r$ planes is $s^{t+1}$ and in each plane there lie $s^t$ vertexes, the fraction of the vertexes lying outside $L_p$ is $(s^{t+1} - s^t)/s^{t+1}$, i.e., $(1-s^{-1})$. The latter expression coincides with the one in Proposition at *p=1*. It should be noted that, in this case, it was not necessary to make the suggestion of induction.

(2) Let plane $L_p$ of dimension *t+1-p* intersects with plane $T_r$ of dimension *t* for some value of *r* ($r = 1, 2, ..., s$). Then the plane of intersection, $L_{p,r}$, has dimension of *t-p*, i.e., its codimension in plane $T_r$ is *p*. Under the suggestion of induction, it follows that the fraction of points M in plane $T_r$, which are lying outside plane $L_{p,r}$, is more than $(1-s^{-p})$. Without the loss of generality, it can be suggested that plane $L_p$ intersects with all planes (4). (Obviously, this suggestion can lead only to the reduction of the number of points lying outside plane $L_p$.) Since planes (4) have no common points and the fraction of the points lying outside plane $L_p$ is not less than $(1-s^{-p})$ for all planes $T_r$, the same result of fraction $(1-s^{-p})$ will be obtained $M_{t+1}$, which proves Proposition.

If the random choice of a sample has uniform probability, then Proposition 1 has an important

**Corollary**. If, among the already-chosen samples, there exist $p$ linearly-independent ones, the probability of the next random sample being linearly-independent (if such sample still exists) equals $1-s^{-p}$. The probability reaches its minimum, $1-0.5^{-p}$, at $s=2$. In particular, in the case of the plane of codimension 1 (hyperplane), not less than 0.5 points of set $M$ (samples) remain outside this hyperplane.

According to Proposition 1, the probability of finding the next linearly-independent sample is a function of the dimension of the already found basis vectors with respect to the full dimension of the sample set. Since the latter parameter is unknown (the evaluation of the full dimension being just the goal of the study), at each step we should choose the probability corresponding to codimension 1. Obviously, this probability value is the minimal of all codimensions, thus the estimation being correct.

Consider now the situation when some probability measure is preset over the set of all possible samples. In the general case (in the framework of Model 2), let us arrange the numbers of genomes corresponding to each coordinate in the order of descending probability of finding a sample with the number of genomes $p(a_{i1}) \geq p(a_{i2}) \geq \ldots p(a_{is})$ ($i=1, 2,\ldots, n$). Next, define the probability of each VGM, $P(T)$, as the product of the probabilities of its coordinates. According to Proposition 1, the number of VGMs lying outside a plane of any codimension is not less than the number of VGMs lying in this plane. However, now their total probability measure may be less than 0.5 and a different estimation is required. For an algorithmic estimation, there is no need to evaluate the total probability value for all codimensions. As it was mentioned above, it is enough to obtain the estimation for codimension 1.

**Proposition 2**. If the probability distribution results in inequalities $p(a_{i1}) \geq p(a_{i2}) \geq \ldots p(a_{is})$, the probability measure of all the vectors of set $M$ lying outside hyperplane $L_1$ of codimension 1 is not less than

$$1 - \frac{1}{(1+\lambda)}, \quad \lambda = \min_{1 \leq i \leq n} (\sum_{2 \leq j \leq s} p(a_{ij}) / p(a_{i1})). \qquad (5)$$

**Proof**. Outside hyperplane $L_1$, there necessarily lies at least one unit vector $e_i$, whose coordinates are all zero and only coordinate $i$ equals 1. Otherwise, if all unit vectors lie in the hyperplane, its dimension coincides with the dimension of the whole space. Consider vector $T$ belonging to the set of genome multiplicities, $M$, and lying in hyperplane $L_1$. The value of the $i^{th}$ coordinate of vector $T$, $a_{iq}$, has probability $p(a_{iq})$, where $q$ is some integer between 1 and $s$. Then, obviously, vector $T - p(a_{iu})/p(a_{iq})e_i \quad (u \neq q)$ also belongs to set $M$, but lies outside plane $L_1$. By definition, its probability is equal to $p(a_{iu})/p(a_{iq})P(T)$. Thus each vector $T$ of set $M$ which lies in plane $L_1$ can be associated with $s$-1 vectors that lie out of this plane. This association is unique in the sense that, for each pair of vectors $T_1 \neq T_2$ $(T_1, T_2 \in L_1)$, equation $T_1 - \alpha_1 e_i = T_2 - \alpha_2 e_i$ is not possible. Furthermore, if $q>1$, also for $u=1$ ratio $p(a_{iu})/p(a_{iq})$ is more than unit on the strength of the suggestion about the ordered probabilities (see above). As a result, the measure of some vectors $T - p(a_{iu})/p(a_{iq})e_i \quad (u \neq q)$ lying outside plane $L_1$ will be larger than the measure of original vector $T$ in plane $L_1$ ($p(a_{iu})/p(a_{iq})P(T) > P(T)$). To estimate the minimum of ratio $p(a_{iu})/p(a_{iq})$, we assume that at coordinate $i$, index $q=1$ for all vectors $T$ lying in plane $L_1$. Then, for index u=2, ratio $p(a_{i2})/p(a_{i1})$ will be the largest of all possible ones, descending order $p(a_{i3})/p(a_{i1}) \geq p(a_{i4})/p(a_{i1}) \geq \ldots$ holding to the last value of $s$. Thus, if the total measure of vectors belonging to set $M$ and lying in plane $L_1$ is equal to $P(L_1)$, it will obey inequality $P(\bar{L}_1) \geq \lambda P(L_1)$, where $P(\bar{L}_1)$ is the full measure for the vectors lying out of the plane, $\lambda = \min_{1 \leq i \leq n}(\sum_{2 \leq j \leq s} p(a_{ij})/p(a_{i1}))$. The ratio of the measure of the vectors lying outside hyperplane $L_1$ to the measure of all the vectors of set $M$ is $Q = P(\bar{L}_1)/(P(L_1) + P(\bar{L}_1))$, which results in $Q > \lambda P(L)/(P(L) + \lambda P(L))$.

.

Thus, $Q > \lambda/(1+\lambda) = 1 - 1/(1+\lambda)$. This estimation coincides with the one obtained in Proposition 1 for the uniform measure multiplicity, $p(a_{i1}) = p(a_{i2}) = \ldots = p(a_{is})$, because in this case $\lambda = s-1$.

In Model 1, in the case of normal distribution of genome multiplicities, the probability of the genome multiplicity value at coordinate $i$ is

$$\frac{1}{\sigma\sqrt{2\pi}} \exp(-\frac{(m_i - \mu_i)^2}{2\sigma^2}),$$

According to (5), $\lambda = \exp(-1/2\sigma_0^2)$, where $\sigma_0 = \min(\sigma_i)$, $i=1, 2,\ldots, n$. For large $\sigma_0$, the value of $\lambda$ approaches 1 (e.g., at $\sigma_0=10$, $\lambda=0.995$).

Next, we evaluate the mean number of samples that have to be taken in order to obtain all the basis vectors of the sample set.

**Proposition 3.** The mean number of samples required to find all $n$ basis vectors is equal to $n/p$, where $p$ is the probability to find a basis sample in one operation.

**Proof.** The probability if finding the $n^{th}$ basis element in the $(n+t)^{th}$ sampling is

$$C_{n+t-1}^{t} p^n (q)^t, \quad q = (1-p)$$

(according to Feller [11], vol. I, VI, 2). Then the mean length of the sampling sequence up to the step of finding all $n$ basis elements is

$$D = \sum_{t=1}^{\infty} (n+t) C_{n+t-1}^{t} p^n (q)^t,$$

or

$$D = p^n \sum_{t=1}^{\infty} \frac{(n+t)(n+t-1)\ldots(n)}{t!} (q)^t = p^n n \sum_{t=1}^{\infty} \frac{(n+t)(n+t-1)\ldots(n+1)}{t!} (q)^t. \quad (6)$$

It is easy to show that

$$\sum_{t=1}^{\infty} \frac{(n+t)(n+t-1)\ldots(n+1)}{t!} (q)^t = (1-q)^{-n-1}.$$

Substituting the latter expression into (6), we obtain $D = p^n n (1-q)^{-n-1} = np^{-1}$, which proves Proposition.

In the case of uniform distribution, considered in Proposition 1, $p = 0.5$ and the mean value of samples is $2n$, while, according to Proposition 2, for any measure used in it, the mean value of samples is $(1+\lambda)n/\lambda$.

The last step in building the above algorithm is formulation the rule of its termination. It follows from the above that, if in $m$ consequent samplings, the spectra are found to be linearly dependent on the already chosen basic set, then, e.g., in the case of uniform distribution of the samples, the probability of all the basis samples having been already found is $2^{-m}$. For example, at $m=4$, it can be claimed, with accuracy of 94%, that there exist no more independent samples.

## 3. Assessing qualitative composition of the medium with respect to known bacteria

On the basis of the results obtained in Section 2, below we propose two different possible algorithms for assessing the qualitative contents of the medium with respect to known bacteria.

*3.1. Algorithm employing the set of basis samples*

The set of basis samples, considered in Section 2 can be directly used for checking the presence of a fixed bacterium in the medium. Indeed, by definition, any sample of the medium can be represented as linear combination of the BS vectors of the basis samples. Any fixed genome present in the medium also represents a sample (although not random) and should be decomposed over the sample basis. If this genome is not present in the medium and, consequently, no basis sample contains it, its decomposition over the sample basis is impossible in view of our fundamental statement formulated in Introduction.

*3.2. Combined algorithm employing the BS of known genomes and the set of basis samples*

Consider, for simplicity, set $B_0$ consisting of BS of all known genomes. At the first stage, for any medium sample, $p_1$, an attempt is made to decompose the sample BS over basis $B_0$. This technique was previously discussed [9] in detail, with regard for different lengths of the squamation fragments and errors of sequencing (see, also [12]). If the BS of sample $p_1$ can be completely decomposed over basis $B_0$, the problem is solved. It should be emphasized that we know not only which genomes are present in the medium, but also their multiplicities, including the error estimation [9]. Suppose now that the medium contains q unknown genomes. This situation will manifest itself by the existence of a residual term in the sample $p_1$ decomposition over basis $B_0$. Formally speaking, this means that q vectors are missing in basis $B_0$ and the consequent samples should be used for supplementing the basis. Without going into details, it should be noted that the problem of choosing linearly independent samples is similar to that considered in Section 2 and the estimate of 0.5 obtained for the probability of choosing a linearly-independent sample (if it still exists) is also true for this case. In this way, the basis set of type ($B_0$, $p_1,..., p_q$) is obtained, and, on the strength of its construction, any sample of the medium is decomposed over this set. If a vector from set $B_0$ is present in this decomposition with non-zero coefficient, the genome corresponding to this vector is present in the sample. However, the sample under consideration may lie in the hyperplane which does not include the BS of the genome to be identified, i.e., the sample may be a vector of "non-common position". It can be shown that, for any fixed genome from set $B_0$, the probability of such event is equal to $1/s$, where $s$, the same as above, is the number of possible multiplicities of a genome in the sample ($s>1$). Thus a few consequent samples will contain the fixed genome or will show that, with great probability, it is not present in the medium.

## 4. Conclusion

The paper presents the theoretical foundations of the algorithm for calculating the number of different genomes in the medium and two algorithms for determining the presence of the fixed genome in the medium under study. All these algorithms are based on repeated sampling (and sequencing of the samples) of metagenomes from the medium, which is a rather time-consuming process. On the other hand, the advantage of the proposed algorithms is the fact that the presence of unknown genomes does not interfere with the calculations. The second algorithm, which uses the data on the already known genomes, is quite economical since the number of the medium samples depends on the number of unknown genomes in this medium.

In the report to follow, we are considering the computational schemes of the proposed algorithms.